\documentclass[prc,twocolumn,amsmath,amssymb,superscriptaddress,nofootinbib,preprintnumbers]{revtex4-1}
\usepackage{url,color,colordvi,epsfig,pstricks}
\usepackage[colorlinks=true, citecolor=blue, filecolor=blue, linkcolor=blue, urlcolor=blue, pdftex]{hyperref}
\usepackage{graphicx,bm,dcolumn}
\usepackage{hyperref,slashed,lineno}
\usepackage{physics}
\usepackage{inputenc}

\hypersetup{
    colorlinks=true, %set true if you want colored links
    linktoc=all, %set to all if you want both sections and subsections linked
    linkcolor=blue, %choose some color if you want links to stand out
    citecolor=blue}

\begin{document}

\preprint{NT@UW-25-7}

\title{Quark Phase Space Distributions in Nuclei}
\author{A.~Nikolakopoulos}\email{anikolak@uw.edu}
\affiliation{Department of Physics, University of Washington, Seattle, WA 98195-1560,USA}
\author{G.~A.~Miller}\email{miller@uw.edu}
\affiliation{Department of Physics, University of Washington, Seattle, WA 98195-1560,USA}
%#######################################################################################

\begin{abstract}
In [PRC 110, 025201], the authors construct a model for nuclear matter which features a quarkyonic phase.
A main feature in this model is that the nucleon occupation is strongly reduced at small momenta.
Somewhat surprisingly, this result is supported in comparison with data for electron scattering from nuclear matter, where a reduction of the cross section consistent with suppression of nucleons with small momenta is seen.
Since nuclear matter data are obtained by extrapolation of electron scattering data on increasingly heavier systems, this feature should manifest at least to some degree in heavy nuclei.
To check if this is plausible we extend the approach of [PRC 110, 025201] to finite nuclei by considering the nuclear Wigner distribution.
We use non-relativistic and relativistic independent particle models to determine the nuclear Wigner distribution, in addition to the local-density approximation (LDA).
Phase-space distributions of quarks are obtained as a convolution of the Wigner distribution with a quark momentum distribution.
We highlight some properties of the Wigner distribution in spherical systems, which can spoil the interpretation of the quark phase-space distribution as occupation numbers in the Fermi sea.
On the other hand, we show that large systems behave essentially like infinite nuclear matter in their interior, and that LDA and full results are quantitatively similar for large A.
We then compute the fraction of baryons that would be in a quarkyonic phase in the same sense as in [PRC 110, 025201] for a set of nuclei with mass $12 \leq A \leq 238$. We find that this fraction systematically tends to a constant at large $A$. It is hence plausible that the suppression seen in the nuclear matter data is a genuine feature, present in large finite nuclei. 
This result is counter-intuitive and we discuss possible electron scattering measurements that could rule out this model.
\end{abstract}

\maketitle

%#######################################################################################
\section{Introduction}

One of the intriguing questions about nuclear physics involves determining the role of quarks and gluons in nuclear physics.
Successful descriptions of nuclear properties make no explicit reference to the underlying degrees of freedom of the QCD Lagrangian. See for example, the reviews \cite{Hagen:2013nca,Barrett:2013nh,Hergert:2015awm}. Yet, observations known as the EMC effect clearly demonstrated that the quark distribution functions of bound nucleons are different from those of free nucleons. See the reviews~\cite{Geesaman95,Norton:2003cb,Hen:2013oha,Hen:2016kwk}.  Thus, determining whether or not quark-gluon effects have an explicit role in nuclear physics remains an important task.

One way approach to quark degrees of freedom in nuclear matter is to consider the possibility of quarkyonic matter~\cite{McLerran:2007qj} in which high baryon density systems consist of a Fermi sphere of quarks surrounded by a momentum shell of baryons. In this picture, the probability of nucleons having small values of momentum is strongly reduced.
Ref.~\cite{McLerran:2018hbz} presented arguments that quarkyonic matter naturally explains the maximum of the density dependence of the speed of sound extracted from analyses of neutron star data~\cite{Fujimoto:2019hxv,Tan:2020ics}.  

Many dynamical models for quarkyonic matter have been proposed~\cite{Jeong:2019lhv,Kojo:2021ugu,Kovensky:2020xif,Cao:2020byn,Duarte:2020xsp,Sen:2020peq,Poberezhnyuk:2023rct}, but it was not clear that those models obtain a state of lowest energy
\cite{Koch:2022act}. That issue was addressed in a new construction for quarkyonic matter-the so-called IdylliQ model of quarkyonic matter~\cite{Fujimoto:2023mzy}. In that model, the quark Pauli principle is responsible for the suppression of nucleons of low momentum.

Until recently, these quarkyonic models were aimed at understanding high-density baryon matter. Ref.~\cite{Koch:2024qnz} considered the relevance of quarkyonic ideas to understanding infinite nuclear matter at normal densities $\approx 0.17 \, \rm fm^{-3}$. This leads to a strong depletion of low momentum nucleon states, which was shown to be consistent with extrapolations of $(e,e')$ quasi-elastic scattering data from finite nuclei to the nuclear matter limit~\cite{Day:1989nw}.

The aim of the present paper is to  extend the analysis of \cite{Koch:2024qnz} to finite systems. In particular we study whether it is plausible that low-momentum states are suppressed in heavy nuclei.
 This is done by replacing the occupation numbers in the Fermi sea by the Wigner distribution of finite nuclei. 
The procedure for this is presented in Sect.~II. Sec.~III shows computations of the quark phase-space density for finite-sized nuclei using both the local density approximation and exact results from mean-field calculations.
Sect.~IV suggests experimental tests of the quarkyonic idea and the paper is summarized in Sect.~V.
\section{Nuclear Wigner distributions}
We extend the analysis of \cite{Koch:2024qnz} to finite systems by considering the Wigner distribution of finite nuclei.
The analysis of Ref.~\cite{Koch:2024qnz} is based on occupation numbers in the Fermi sea for nuclear matter.
The baryon density in non-interacting nuclear matter is
\begin{equation}
\rho = 2 \int \frac{\mathrm{d}^3 \mathbf{k}}{(2\pi)^3}~f_p (k) + f_n (k) = 4 \int \frac{\mathrm{d}^3 \mathbf{k}}{(2\pi)^3}~f_N(k)
\end{equation}
where $f_p$ / $f_n$ are the proton/neutron occupation numbers, and $f_N$ the nucleon occupation number in isospin-symmetric matter.
The phase-space density for quarks is then obtained as the convolution~\cite{Fujimoto:2023mzy}
\begin{equation}
\label{eq:fq_occ}
f_q(p) = \int \frac{\mathrm{d}^3\mathbf{k}}{(2\pi)^3} \phi( \lvert \mathbf{p} - \mathbf{k}/N_c \rvert ) \left[ f_p (k) + f_n (k) \right],
\end{equation}
with $\phi(\mathbf{p})$ the quark momentum distribution of the nucleon.
This equation is a statement that the quark distribution is a   product of the probability for a quark to exist in a nucleon of a given momentum with the probability for the nucleon to have that momentum.

To extend this to finite-sized nuclei, we consider the Wigner quasi-probability distribution~\cite{PhysRev.40.749} of nuclei given by
\begin{equation}
\label{eq:def_Wign}
W(\mathbf{r},\mathbf{p}) = \int \mathrm{d}^3\mathbf{s} e^{i\mathbf{p}\cdot\mathbf{s}} \rho(\mathbf{r} - \mathbf{s}/2, \mathbf{r} + \mathbf{s}/2),
\end{equation}
where $\rho(\mathbf{r},\mathbf{r}^\prime)$ is the one-body density matrix. 
We then obtain the phase-space density of quarks as
\begin{equation}
\label{eq:F_Q}
W_q\left( \mathbf{r},\mathbf{p} \right) = \int \frac{\mathrm{d}^3\mathbf{k}}{(2\pi)^3} \phi( \lvert \mathbf{p} - \mathbf{k}/N_c \rvert) W(\mathbf{r},\mathbf{k}).
\end{equation}
In non-interacting nuclear matter, Eqs.~(\ref{eq:fq_occ}) and (\ref{eq:F_Q}) are equivalent.

We consider two approaches for the one-body density matrix using the following. 
First, a straightforward extension of the nuclear matter case in the local-density approximation (LDA).
Then, calculations in the independent particle model using relativistic mean-field (RMF) wavefunctions, and wavefunctions obtained from non-relativistic Hartree-Fock (Bogoliubov) calculations.

\subsection{Nuclear Matter and Local Density Approximation}
\label{sec:nuclear_matter}
The one-body density matrix for nuclear matter is given by
\begin{equation}
\rho(\mathbf{r}, \mathbf{r}^\prime) = 2  \int_0^{k_F} \frac{\mathrm{d}^3\mathbf{k}}{(2\pi)^3} e^{-i\mathbf{k}\cdot\left(\mathbf{r} - \mathbf{r}^\prime\right)}
\end{equation}
where $2$ is the spin degeneracy, and we consider a single isospin state. 
The one-body density matrix can thus be written
\begin{align}
\rho(\mathbf{r}+\mathbf{s}/2, \mathbf{r} - \mathbf{s}/2) &= \frac{1}{\pi^2} \int_0^{k_F} \mathrm{d} k  k^2 j_{0}(ks) \nonumber \\
 &= \frac{3}{k_F s} j_1(k_F s) \rho(\mathbf{r},\mathbf{r}),
\label{eq:OBDM_RFG}
\end{align}
where $j_n(k)$ are spherical Bessel functions and $\rho(\mathbf{r},\mathbf{r})={k_F^3\over 3\pi^2}$.
The Wigner distribution for nuclear matter is then obtained as
\begin{align}
\label{eq:Wrp_matter}
W(\mathbf{r},\mathbf{p}) &= \int \mathrm{d} \mathbf{s} e^{-i\mathbf{p}\cdot\mathbf{s}} \rho(\mathbf{r}+\mathbf{s}/2,\mathbf{r}-\mathbf{s}/2)  \\
&= \rho(\mathbf{r},\mathbf{r}) \frac{ 6 \pi^2}{k_F} \int \mathrm{d}s~ J_{1/2}(ps) J_{3/2}(s k_F) = 2 f(p),
\end{align}
where $J_n(x)$ are the cylindrical Bessel functions and~\cite{NIST:DLMF_10_22_63} 
\begin{equation}
f(p) =
\begin{cases}
1 & p < k_F \\
0 & p > k_F
\end{cases}
\end{equation}
Hence, for non-interacting nuclear matter, the Wigner distribution is equivalent to the occupation number.

A straightforward way to treat nuclei is then to use a local density approximation.
Given the radial density, the local Fermi momentum is defined as
\begin{equation}
k_F(r) = \left(3\pi^2 \rho(\mathbf{r},\mathbf{r}) \right)^{-1/3}.
\end{equation}
The $r$-dependence is introduced in the one-body density matrix of Eq.~(\ref{eq:OBDM_RFG}) through the $r$-dependent local Fermi momentum, and we have the Wigner distribution
\begin{equation}
W_{LDA}(\mathbf{r}, \mathbf{p})/2 =  f(r,p) 
\begin{cases}
1 & p < k_F(r) \\
0 & p > k_F(r)
\end{cases}
\end{equation}

\subsection{Wigner distribution in spherical nuclei}
\label{sec:RMF}
We compute the Wigner distribution for finite nuclei, using a mean-field description of the nucleus.
Details of the calculation, and derivation of the properties of the Wigner distribution are given in App.~\ref{app:Wigner_RMF}.
In the mean-field approximation, the ground state is a Slater determinant of single-particle states $\psi_{n,\kappa}^{m_J}$, labeled by principal quantum number $n$, relativistic angular momentum $\kappa$ and projection of total angular momentum $m_J$.
The contribution to the Wigner distribution for each occupied single-particle state is
\begin{equation}
W_{n,\kappa}^{m_J} (\mathbf{r},\mathbf{p}) = \int \mathrm{d}\mathbf{s} e^{i\mathbf{s}\cdot\mathbf{p}} \left[ \psi^{m_J}_{n,\kappa} (\mathbf{r}-\mathbf{s}/2)\right]^\dagger \psi_{n,\kappa}^{m_J}(\mathbf{r}+\mathbf{s}/2).
\end{equation}
And the full Wigner distribution is given by
\begin{equation}
W(\mathbf{r},\mathbf{p}) = \sum_{n,\kappa} W_{n,\kappa}(\mathbf{r},\mathbf{p}) = \sum_{n,\kappa,m_j} W_{n,\kappa}^{m_J}
\end{equation}
For filled shells, the sum over $m_J$ can be performed immediately and one has
\begin{equation}
\label{eq:Wnk_rp}
W_{n,\kappa}(\mathbf{r},\mathbf{p}) = \int \mathrm{d}\mathbf{s} e^{i\mathbf{p}\cdot\mathbf{s}} \rho_{n,\kappa}(\mathbf{r}-\mathbf{s}/2, \mathbf{r}+\mathbf{s}/2),
\end{equation}
with $\rho_{n,\kappa}(\mathbf{r},\mathbf{r}^\prime)$ given in Eq.~(\ref{eq:rho_kappa}).
Here we highlight a number of differences with respect to the LDA considered above.

Because of the rotational symmetry of the system, instead of translational invariance, the one-body density matrix $\rho(\mathbf{r} - \mathbf{s}, \mathbf{r}+\mathbf{s})$ depends on the angle $\cos\theta_{r,s} = \hat{\mathbf{r}}\cdot\hat{\mathbf{s}}$. The Wigner distribution thus depends on the relative angle $\cos\theta_{r,p} = \hat{\mathbf{r}}\cdot\hat{\mathbf{p}}$.
This angular dependence is fairly weak, and is irrelevant for our main results.
Hence, for purposes of presentation in the following, we introduce the angle-averaged distribution
\begin{equation}
w(r,p) \equiv \frac{1}{2} \int \mathrm{d}\cos\theta_{r,p} W(\mathbf{r},\mathbf{p}).
\end{equation}

The momentum is not a good quantum number in a self-bound %spherical 
system so that $w(r,p)$ is required to take on negative values for some part of the phase-space.
Furthermore, while in the LDA $W(\mathbf{r},\mathbf{p}) \leq 2$, one instead derives the bound $\lvert w_{n,\kappa}^{m_J}(r,p) \rvert \leq 8$ in a spherical system. As such $\lvert w_{n,\kappa}(r,p) \rvert \leq (2J+1) 8$ for each shell. 
We show that this bound is almost saturated for small $r$ and $k$ in Appendix~\ref{app:Wigner_RMF}.
Indeed, for $r=0$ one finds
\begin{align}
w_{n,\kappa}(0,p) = 8(-1)^l\tilde{\rho}^{S}_{n,\kappa}(p/2),
\end{align}
where $\tilde{\rho}_{n,\kappa}^{S}(p)$ is the Fourier transform of the scalar density of the shell.
For small values of $p$, the scalar density almost saturates the bound $\rho^{S}_{n,\kappa}(0) \lesssim (2J+1)$ in a relativistic mean field. It exactly satisfies this bound in a non-relativistic calculation.
Since we fill shells that contribute with alternating sign $(-1)^l$ to the total Wigner distribution, the maximal value of $\lvert w(r,p) \rvert$ turns out to be much smaller than the strict bound of $8A$.

\begin{figure*}
\includegraphics[width=0.32\textwidth]{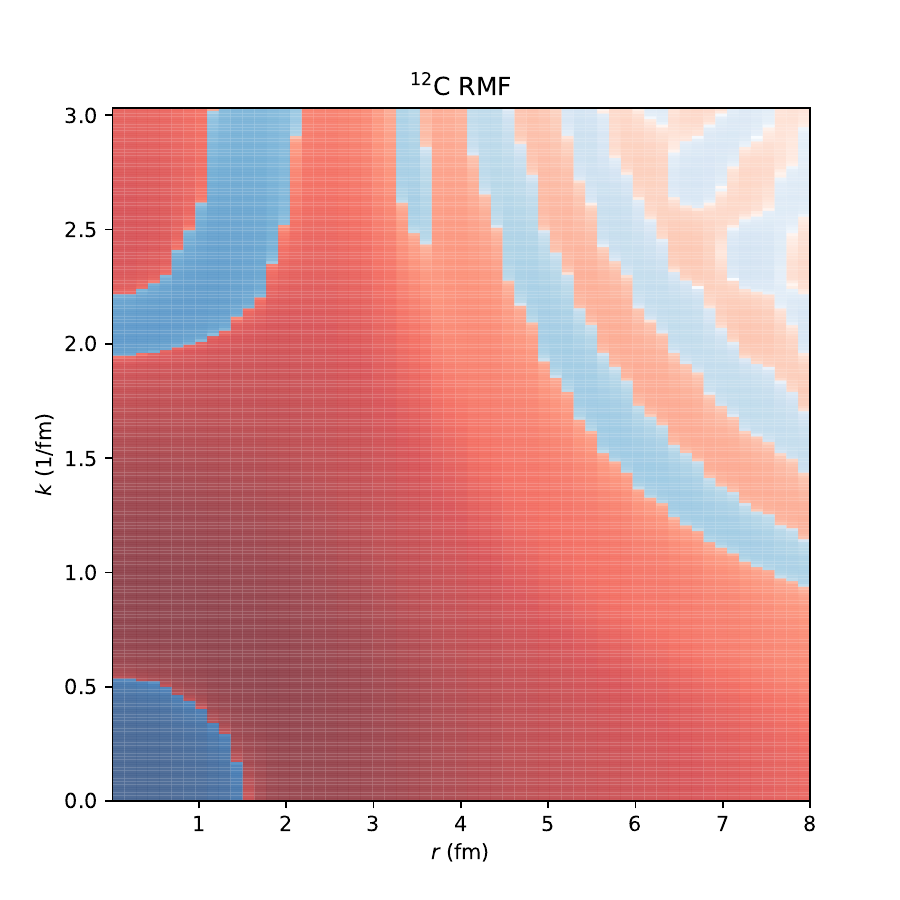}
\includegraphics[width=0.32\textwidth]{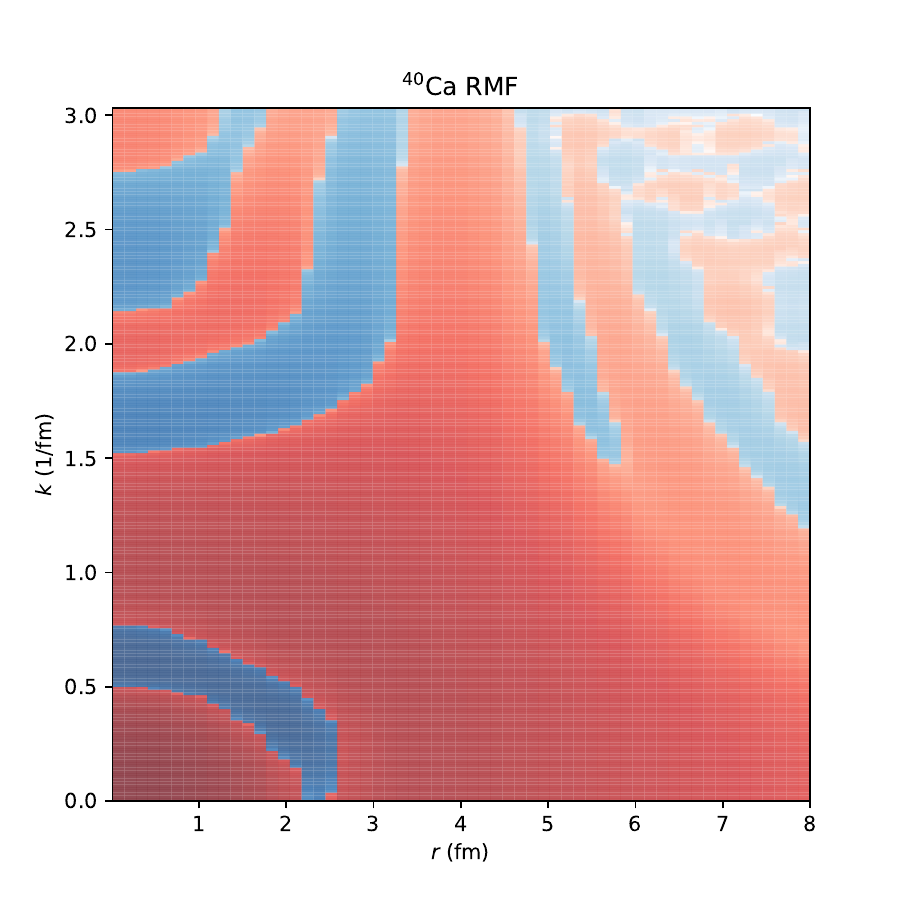}
\includegraphics[width=0.32\textwidth]{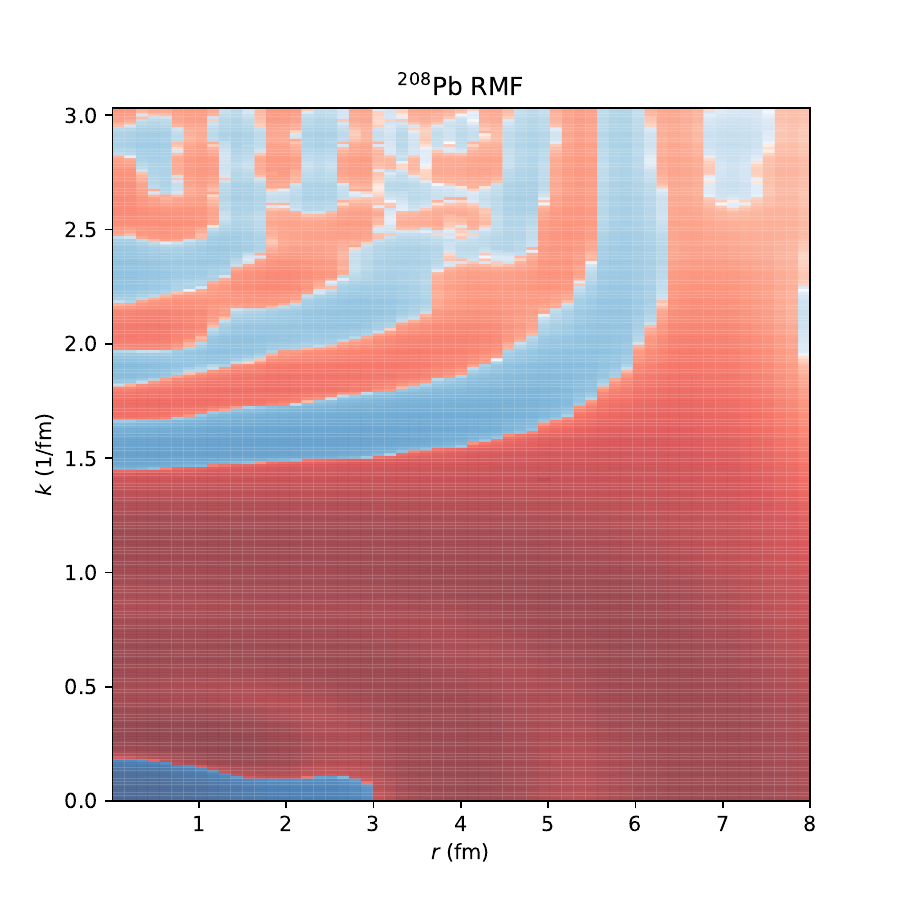}

\includegraphics[width=0.32\textwidth]{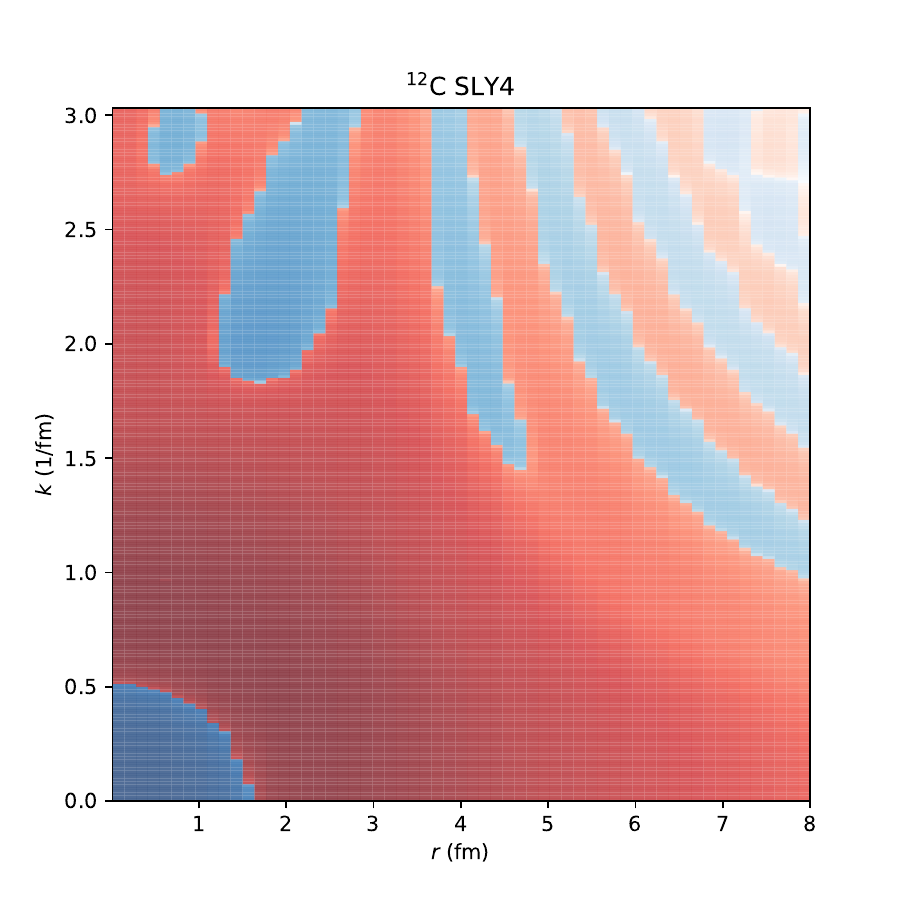}
\includegraphics[width=0.32\textwidth]{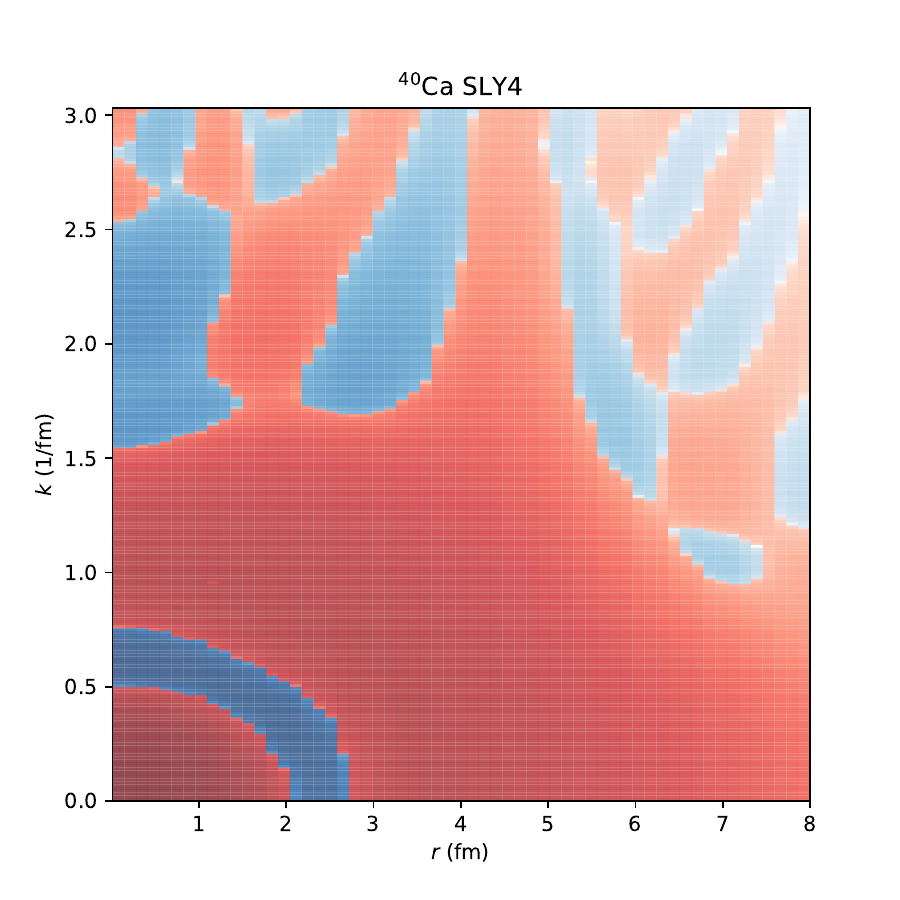}
\includegraphics[width=0.32\textwidth]{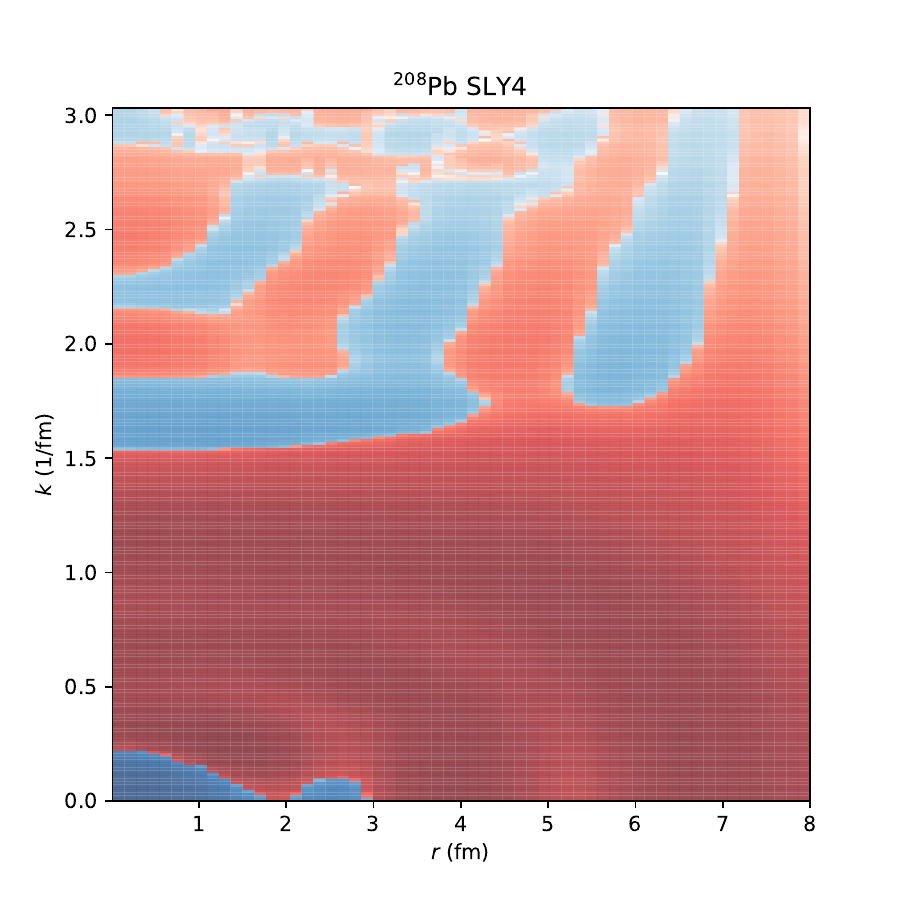}
\caption{Angle averaged Wigner distribution $w(r,k)$ for ${}^{12}$C (left) , ${}^{40}$Ca (middle), and ${}^{208}$Pb (right) on logarithmic scale. 
Blue and red correspond to negative and positive values of $w(r,k)$ respectively. Top row is computed with RMF wavefunctions, bottom row are HF calculations with the SLY4 force.}
\label{fig:2d_wrp}
\end{figure*}

For this work we use Wigner distributions from Hartree-Fock (HF) calculations~\cite{Bennaceur:2005mx}, and relativistic mean field (RMF) calculations~\cite{POSCHL199775,Horowitz1991}.
We use the NL-SH parameter set of Ref.~\cite{SHARMA1993377} for the RMF calculations, and the SLY4 functional of Ref.~\cite{CHABANAT1998231} for the HF calculations. As will become evident the exact details of the functionals do not affect our main results. 
The Wigner distributions for ${}^{12}$C, ${}^{40}$Ca, and ${}^{208}$Pb are shown in Fig.~\ref{fig:2d_wrp}.
The RMF and HF results are found to be quantitatively similar. The results for carbon and calcium can also be compared to the independent particle model results of Ref.~\cite{Cosyn:2021ber}, where the same structure is found.
One can understand this behavior of $w(r,p)$ at small radii and momenta as follows. The limiting behavior of the Wigner distribution for small $r$ and small $k$, given in Eqs.~(\ref{eq:wign_r0}-\ref{eq:wign_p0}), implies that $w_\kappa$ in this region is mostly determined by the Fourier transform of the density and momentum distribution of the shell respectively. That is: the normalization and radius of the densities.
The shells then contribute with sign $(-1)^l$ to the total Wigner distribution. As such, the similarity between the different calculations is a result of the angular momentum structure of the nuclear shell model. In ${}^{12}$C for example, 2 nucleons in the $s_{1/2}$ and 4 nucleons in the $p_{3/2}$-shell combine to form the negative island in the center with $w(0,0)\approx -16$.

\subsection{Quark phase-space distribution}
Given the momentum distribution for quarks in the nucleon $\phi(\mathbf{k})$, normalized such that
\begin{equation}
\label{eq:norma_phi_q}
1 = \int \frac{\mathrm{d}\mathbf{k}}{(2\pi)^3} \phi(\mathbf{k}),
\end{equation}
we compute the phase-space distribution of quarks in the nucleus as the convolution
\begin{equation}
W_q(\mathbf{r},\mathbf{k}) =  \int \frac{\mathrm{d}\mathbf{p}}{(2\pi)^3} \phi\left( \mathbf{k} - \mathbf{p}/N_c \right) W(\mathbf{r},\mathbf{p}),
\label{18}
\end{equation}
where $N_c$ is the number of colors.
It is useful to write this convolution as
\begin{equation}
W_q(\mathbf{r},\mathbf{k}) = \int \mathrm{d}\mathbf{r}^\prime e^{-i\mathbf{k}\cdot\mathbf{r}^\prime} \tilde{\phi}(\mathbf{r}^\prime) \tilde{W}(\mathbf{r},\mathbf{r}^\prime/N_c),
\end{equation}
where the tildes denote the Fourier transform.
Clearly $\tilde{W}(\mathbf{r},\mathbf{r}^\prime) = \rho(\mathbf{r}-\mathbf{r}^\prime/2, \mathbf{r} +\mathbf{r}^\prime/2)$ and $\tilde{\phi}(\mathbf{r}) = \tilde{\phi}(r)$ such that
\begin{equation}
\label{eq:F_q_rspace}
W_q(\mathbf{r},\mathbf{k}) = \int \mathrm{d}\mathbf{s} e^{-i\mathbf{k}\cdot\mathbf{s}} \tilde{\phi}(s) \rho(\mathbf{r} - \frac{\mathbf{s}}{2N_c}, \mathbf{r} + \frac{\mathbf{s}}{2N_c}).
\end{equation}
Since Eq.~(\ref{eq:F_q_rspace}) has the same form as Eq.~(\ref{eq:Wnk_rp}), $W_q$ can be computed in the same way as the Wigner distribution. 
This distribution depends (weakly) on the relative angle $\hat{\mathbf{k}}\cdot\hat{\mathbf{r}}$, and it is more convenient to consider the angle-averaged distribution
\begin{equation}
w_q(r,k) \equiv \frac{1}{2}\int \mathrm{d}\cos\theta_{r,k} W_q(\mathbf{r},\mathbf{k}).
\end{equation}
This is an accurate approximation for the results obtained in this work. 
Indeed, $\tilde{\phi}(s)$ is only appreciable at scales comparable to the spatial extent of the  nucleon $R_N$, much smaller than the size of the nucleus $R_A$. The ratio $s/(6 R_A)\sim R_N/(6R_A)\ll 1$. Moreover, for our main results we will only require $W_q$ at small momenta $k \lesssim 0.5~\mathrm{fm}^{-1}$.
Thus, the $l=0$ partial wave of $e^{-i{\bf k}\cdot{\bf s}}$ is by far the most important so that
it is accurate to consider the angle-averaged distribution\footnote{Note that the angular distribution is symmetric, so the first corrections are $\sim (kR_N\cos\theta)^2$. See Taylor expansion of Eq.~(\ref{eq:W_full}).}
.

We follow Refs.~\cite{Fujimoto:2023mzy,Koch:2022act}, and use a simple model for the quark momentum distribution
\begin{equation}
\label{eq:quark_momentumdist}
\phi(\mathbf{k}) = \frac{2\pi^2}{\Lambda^2} \frac{e^{-\lvert\mathbf{k} \rvert/\Lambda}}{\lvert \mathbf{k} \rvert}, \quad \tilde{\phi}(s) = \frac{1}{1+(s\Lambda)^2}.
\end{equation}
With this form for the momentum distribution, we have the following relation between the quark and nucleon Wigner distribution,
\begin{equation}
\label{eq:diff_wq_wN}
\left( -\nabla^2_{\mathbf{k}} + \frac{1}{\Lambda^2} \right)W_q(\mathbf{r},\mathbf{k}) = \frac{N_c^3}{\Lambda^2} W(\mathbf{r},N_c\mathbf{k}).
\end{equation}

We take the proton-neutron asymmetry into account is a simple way.
We define $u$ and $d$ quark densities from the proton/neutron densities as
\begin{equation}
\label{eq:up_down_densities}
\rho_u \equiv \frac{2}{3}\rho_p + \frac{1}{3}\rho_n,\quad \rho_d \equiv \frac{2}{3}\rho_n + \frac{1}{3}\rho_p, 
\end{equation}
and denote the equivalent linear combinations of other quantities with the same subscripts.

\section{Results}
We proceed as in Ref.~\cite{Koch:2024qnz}, and identify $w_{q}(r,p)/2 = f_q(r,p)$ as occupation numbers in the quark Fermi sea. This considers a single isospin state, and the factor two accounts for the spin degeneracy as in Eq.~\ref{eq:Wrp_matter}. In this case, the Pauli principle would forbid $f_q(r,p) \geq 1$.
Our main goal at present is not to strictly enforce this constraint, but instead study the $A$-dependence of the quark occupation and violation of this bound.

\subsection{Local Density Approximation}
\begin{figure}
\includegraphics[width=0.45\textwidth]{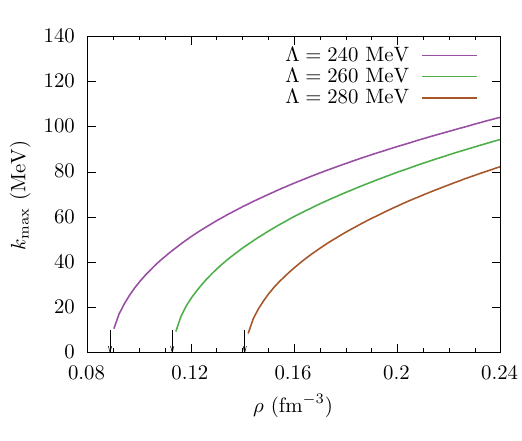}
\caption{The value of $k_{max}$, below which the quark occupation number exceeds 1, as function of the density in symmetric nuclear matter for different values of $\Lambda$. The critical density for each $\Lambda$ is indicated by the arrows.}
\label{fig:rho_p_crit}
\end{figure}
From Eqs.~(\ref{eq:quark_momentumdist}),~(\ref{eq:F_q_rspace}),~and~(\ref{eq:OBDM_RFG}), one sees that in the LDA $f_q(r,k)$ is a monotonically decreasing function of $k$ with maximum at $k=0$.
As such, when $f_q(r,k=0) > 1$, there exists some part of the phase space where $f_q \geq 1$.
Since the quark occupation numbers $f_q(r,k)$ at fixed $r$ are fully determined by the local density, i.e. $k_F(r)$, there is a critical density for which $f_q \geq 1$.
This density is determined in terms of the local Fermi momentum as
\begin{equation}
f_q(r, k=0) = 1 = \frac{N_c}{\Lambda^2} \int_0^{k_F(r)} \mathrm{d}k~ke^{-\frac{k}{\Lambda N_c}}.
\end{equation}
The critical value of $k_F$ is then given by
\begin{equation}
\frac{k_F}{N_c\Lambda} = -W_{-1}\left( \frac{1 - 1/N_c^3}{e} \right) -1,
\end{equation}
where $W_{-1}$ is the lower branch of the Lambert $W$ function, where a positive solution is found. For $N_c = 3$ one finds $k_F = 0.90\Lambda$.
This critical density is larger than the first-order estimate $k_F = N_c\Lambda/\sqrt{2}$ given in Ref.~\cite{Koch:2024qnz}.
 Numerically, for reasonable values of $\Lambda$, the critical density is comparable to those determined in Ref.~\cite{Koch:2022act}. %[Jerry Larry paper].
The region of phase-space in which $f_q > 1$ in the LDA is then determined by a maximum momentum for which $f_q(r,k_{max}) = 1$. The quark occupation exceeds the limit set by the Pauli principle up to this momentum. The critical densities and values of $k_{max}$ are shown in Fig.~\ref{fig:rho_p_crit} for different values of $\Lambda$ for symmetric nuclear matter.

\begin{figure*}
\includegraphics[width=0.99\textwidth]{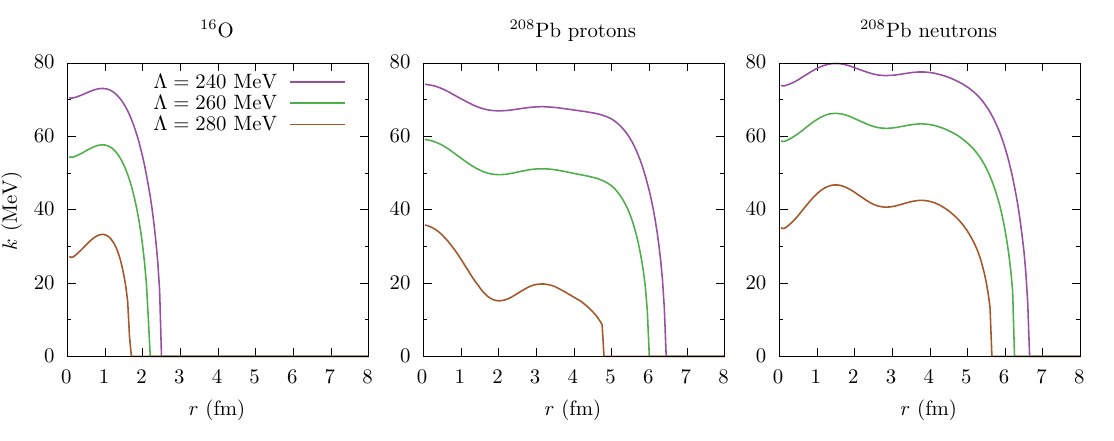}
\caption{Region of phase space where $f_q(r,k) > 1$ in the LDA correspond to the regions below (smaller values of $r$ and $k$) the colored curves. Results are shown for ${}^{16}$O and for the proton/neutron density in ${}^{208}$Pb. Densities are obtained from HF calculations using the SLY4 Skyrme force.  }
\label{fig:contours}
\end{figure*}
These critical densities are comparable to nuclear saturation density, and the central densities of heavy nuclei can exceed the critical values.
In light systems, locally the density may far exceed nuclear saturation density. Thus in the LDA, one obtains a region of phase space in which this bound is exceeded even for light systems.
This argument relies on the LDA however, which should not be realistic for light systems. 
In Fig.~\ref{fig:contours} we show the region of phase-space in which $f_q(r,k)> 1$ for different values of $\Lambda$.
One sees, that while indeed in $^{16}$O there is a pocket of phase-space where the Pauli principle is violated, this region is much smaller than in ${}^{208}$Pb.
We can compute the number of baryons in regions that violate the bound $f_q > 1$ as an integral over the Wigner distribution. As a measure of violation of the quark Pauli principle one can then consider the probability
\begin{equation}
P_{f_q>1} \equiv \frac{1}{A} \int_{f_q > 1} \mathrm{d}\mathbf{r} \frac{\mathrm{d}\mathbf{k}}{(2\pi)^3} w_q(\mathbf{r},\mathbf{p}).
\label{Pq}
\end{equation} 

\subsection{Finite sized nuclei}
\begin{figure}
\includegraphics[width=0.48\textwidth]{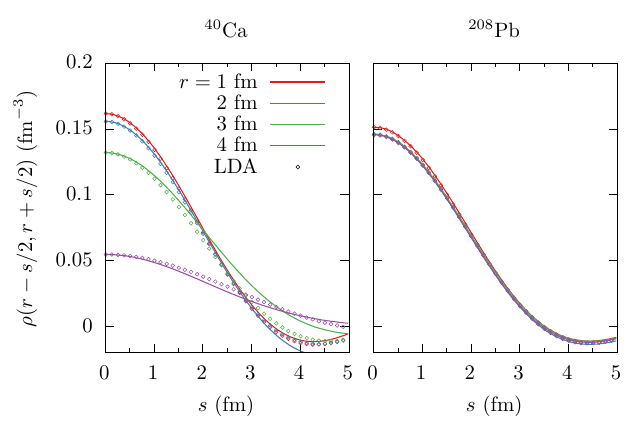} \\
\includegraphics[width=0.48\textwidth]{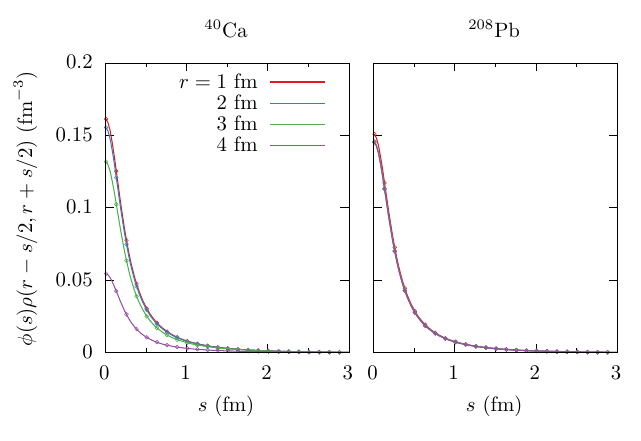}
\caption{Top panels: angle-averaged one-body density matrices for calcium (left) and lead (right), compared to the result obtained in the local-density approximation.
Bottom panels: the same but weighted with the Fourier transform of the quark momentum distribution.}
\label{fig:OBDM}
\end{figure}
The Wigner distribution for finite nuclei differs significantly from those in the LDA. Thus, the reader might question  whether $f_q$ can be interpreted as the occupation probability of quarks in a Fermi gas type system and thus find it  unclear whether the bound on $f_q$ set by the Pauli principle should apply at all  in this case.
By extension this calls into question the applicability of the LDA analysis above.
In the following we will take the bound $f_q >1$ as an indication that quarkyonic degrees of freedom might be relevant as done in the LDA, in order to study the $A$-dependence of a possible quarkyonic phase in nuclei.

\begin{figure*}
\includegraphics[width=0.3\textwidth]{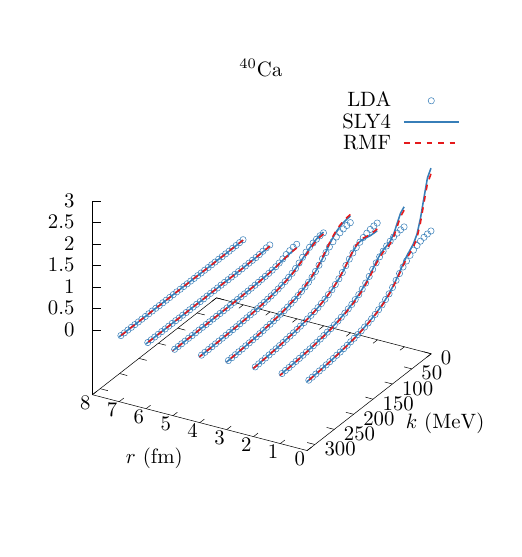}
\includegraphics[width=0.3\textwidth]{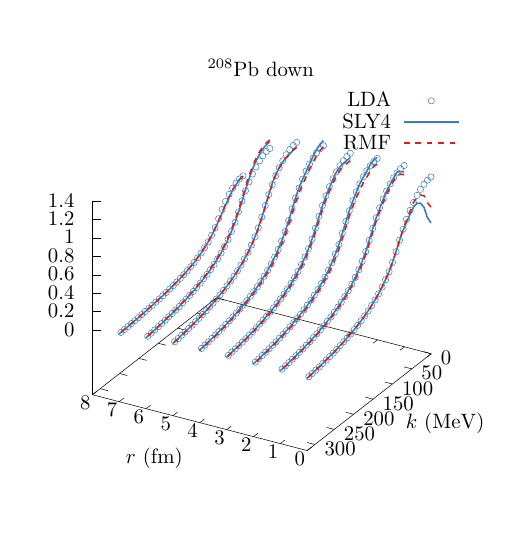} 
\includegraphics[width=0.3\textwidth]{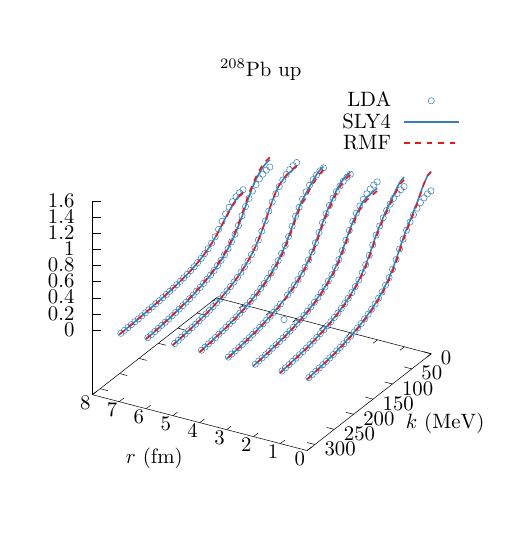} \\
\caption{Quark occupation numbers $f_q = w_q/2$, obtained from the RMF and HF wavefunctions, compared to the LDA results using the HF nuclear density. }
\label{fig:f_q_RMF_LDA}
\end{figure*}
This can be motivated for large systems that (locally) resemble saturated nuclear matter.
Put more precisely, given Eq.~(\ref{eq:F_q_rspace}), the nucleus resembles nuclear matter for our purposes if the one-body density matrix resembles the nuclear matter result of Eq.~(\ref{eq:OBDM_RFG}).
This is indeed the case for large systems.
First, note that Eq.~(\ref{eq:OBDM_RFG}) is the zeroth-order term in the expansion of the one-body density matrix of Ref.~\cite{Negele:1972zp}. Keeping only this  term is a good approximation that becomes more and more accurate as the value of $A$ increases.
Moreover for heavy nuclei, the (isospin summed) density is approximately constant up to the nuclear radius, so that the one-body density matrix at small separation $s$ resembles that of nuclear matter with a global, rather than local Fermi momentum.
This is shown explicitly in Fig.~\ref{fig:OBDM}, where the one-body density matrices for ${}^{40}$Ca and ${}^{208}$Pb, averaged over the relative angle $\cos\theta_{r,s}$ are shown.
It is clear that the LDA provides an excellent approximation at small separation $s$ for both calcium and lead. 
For lead, as explained, one sees that the LDA is excellent up to large separation $s/2 \lesssim 8 \,\mathrm{fm}\gg R_N$, and the one-body density matrix can be reproduced up to large radii with a global Fermi momentum.
Note further that the quark phase-space density of Eq.~(\ref{eq:F_q_rspace}) probes the density matrix at only at small $s \sim R_N$. Hence, the LDA will provide similar results to the full calculation even for calcium.
This can be seen in the bottom panels of Fig.~\ref{fig:OBDM}, where we show the one-body density matrix weighed with the quark momentum distribution for $\Lambda = 260~\mathrm{MeV}$.
As the large $s$ region is strongly suppressed, the large $k$ dependence of $w_q(r,k)$ is approximately a universal function, weighted with the nuclear density. This function is determined by the small-$s$ dependence, i.e. the local Fermi momentum.
Only for small $k$, fluctuations around the LDA result due to the high-$s$ tail of the one-body density matrix are present.
The similarity between $f_q(r,k)$ obtained from the full calculation and in the LDA with the same nuclear density is shown explicitly in Fig.~\ref{fig:f_q_RMF_LDA}.

\subsection{A-dependence of the quarkyonic regime}
\begin{figure*}
\includegraphics[width=0.3\textwidth]{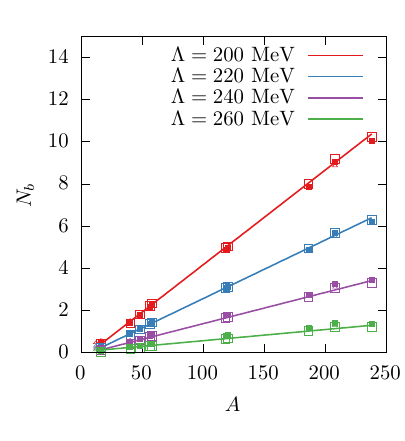}
\includegraphics[width=0.3\textwidth]{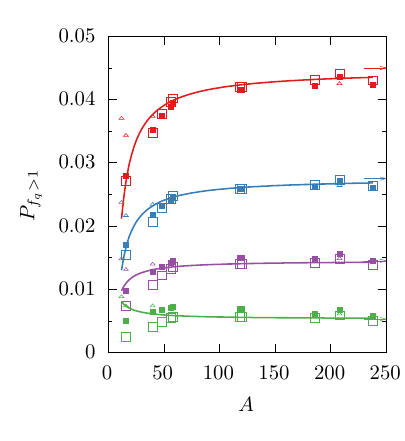} 
\includegraphics[width=0.3\textwidth]{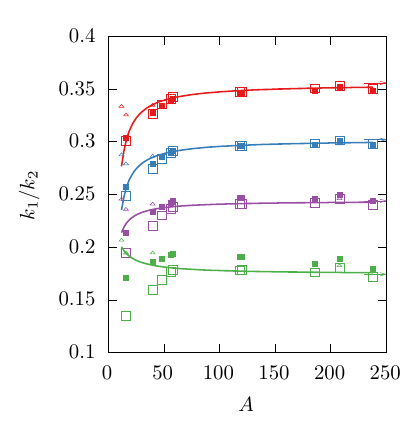} \\
\caption{Number of baryons for which $f_q(r,k) > 1$. Left panel shows the absolute $N_q = N_u + N_d$ as function of $A$ for different values of $\Lambda$. The LDA results are represented by empty squares. The result computed from the full Wigner distribution are shown as filled squares (HFB) and triangles (RMF). The lines are the linear best fit to the LDA results for $A > 50$.
The middle panel shows the ratio $P_{f_q>1}=N_q/A$. The arrows show the asymptotic values for nuclear matter obtained from the fit. Finally, the right-hand panel shows $\left(N_q/A\right)^{1/3}$, which in the nuclear matter limit is the ratio of the 'hole' in the Fermi sea with respect to the equivalent Fermi momentum (see text).}
\label{fig:N_bar}
\end{figure*}
We now consider the condition $f_q(r,k) > 1$ as an indication of quarkyonic degrees of freedom being oversaturated.
The caveat to this assumption have been discussed in the previous section. 
As seen in the previous section, for reasonable choices of $\Lambda$, $f_q(r,k) > 1$ in part of the phase-space.
From LDA results, it is clear that this region of phase space is relatively small for light systems, and it grows for large systems which resemble saturated nuclear matter at least in their cores for $r$ smaller than the nuclear radius.
We take the probability $P_{f_q>1}$ of Eq.~(\ref{Pq}) to indicate the importance of the quarkyonic regime.
To study the $A$-dependence, we consider a set of nuclei, 
${}^{40}$Ca, ${}^{48}$Ca, ${}^{56}$Fe, ${}^{58}$Ni, ${}^{16}$O ,  ${}^{208}$Pb, ${}^{118}$Sn, ${}^{120}$Sn, ${}^{238}$U, ${}^{186}$W.
For the nuclei without completely filled shells we use Hartree-Fock Bogoliubov (HFB) calculations using HFBRAD~\cite{Bennaceur:2005mx}, again with the SLY4 functional~\cite{CHABANAT1998231} as explained in Appendix~\ref{app:Wigner_RMF}. 
We compute $A P_{f_q>1}=N_q$ both from the full one-body matrix, and in the LDA.
We also include RMF calculations for ${}^{12}$C, ${}^{16}$O, ${}^{40}$Ca, and ${}^{208}$Pb using the NL-SH parameterset of Ref.~\cite{SHARMA1993377}.
We take into account the N/Z asymmetry by computing $N_{u/d}$ using the up/down quark densities defined in Eq.~(\ref{eq:up_down_densities}), and present $N_q = N_u + N_d$.

Results are shown in Fig.~\ref{fig:N_bar}.
The $A$-dependence of $N_q$ can be fit by a linear function. We show the linear fit to the points computed in the LDA for $A > 50$.
These fit results reproduce the points with smaller $A$ well for $\Lambda > 240$.
The linear relation is expected. In a simplified model for the nucleus of a sphere with constant density this relation would be exact.
The results of the calculation with the full one-body density matrix using HFB wavefunctions are shown by solid squares. These are compatible with the LDA, exhibiting the same linear trend, showing the robustness of this result.
The linear relation means that the probability $P_{f_q > 1} = N_b/A$ tends to a fixed value in the nuclear matter limit.
This is shown in the middle panel of Fig.~\ref{fig:N_bar}. For $A > 50$, the ratio plateaus and quickly tends to the nuclear matter limit.
It is hence feasible that the experimental procedure to determine the cross section for nuclear matter, i.e. an extrapolation to large $A$, could recover this trend.
For large $\Lambda$, the fit can be extrapolated to $A < 50$. In this case it is seen that $P_{f_q > 1}$ decreases sharply for small $A$, indicating a possible smaller effect for light systems.
However, the results for small $A$ are not as robust. One sees that the RMF calculations are incompatible with the HFB for small $A$, but are compatible for calcium and lead.

We can connect this baryon number with the 'hole' in the nucleon momentum distribution introduced in the model of Ref.~\cite{Koch:2024qnz}.
This model constructs a minimum of the energy that satisfies the constraint $f_q < 1$, by maximally filling up quark states in the Fermi gas up to some momentum $k_q$.
The baryon $N_q$ we compute can be identified with this the number of quarks in this region $N_q = 4\pi k_q^3/3$.
Eq.~(\ref{eq:diff_wq_wN}) then implies that the nucleon occupation is strongly reduced, it is $1/N_c^3$ up to momenta $k_{hole} = N_c\lvert \mathbf{k}_q\rvert$.
In a Fermi gas then, $N_B/A = \left(k_{hole}/k_{F}\right)^3$\footnote{$k_F$ here is the maximum occupied momentum in the model of Ref.~\cite{Koch:2024qnz}, it is slightly larger than the Fermi momentum in normal nuclear matter since momentum states are pushed to larger momentum to compensate for the hole at small momenta}.
The right-hand panel of Fig.~\ref{fig:N_bar} shows the ratio $k_{hole}/k_F$.
The size of the 'hole' is of the order of $20\% - 30 \%$, comparable to the nuclear matter results of Ref.~\cite{Koch:2024qnz}.

%\textcolor{red}{Another note, since the ratio $k_1/k_2$ is dimensionless, and determined by a transcendental equation involving only dimensionless momenta $k/\Lambda$, the relation between this ratio and $\Lambda$ at fixed $k_f$ is linear, which is exactly what is found. The value of $\Lambda$ where the ratio goes to zero is given by the critical density from before $k_f = 0.9\Lambda_c$, we find $\Lambda_c  \approx 300$ MeV as this point, finding a Fermi momentum of $\approx 270 MeV$, which is compatible with the (local) Fermi momentum for the neutrons found in the large nuclei we consider. }

\section{Experimental tests}
Previous work Ref.~\cite{Koch:2024qnz} showed that the suppression of low-momentum nucleons due to quarkyonic matter was not ruled out by $(e,e')$ data for nuclear matter in kinematic regions around the quasielastic peak. 
These  data  are is obtained by extrapolation of $(e,e')$ measurements on heavy nuclei to large $A$~\cite{Day:1989nw}. 
For this suppression  to be a genuine feature then, the low-momentum suppression seen for nuclear matter should be present in the data used in the extrapolation.
Our calculations for finite nuclei show that the size of the momentum region that possibly gets suppressed due to the quark Pauli principle indeed systematically tends to the value for constant density with increasing $A$.
It is hence plausible that this effect can survive the extrapolation procedure to large $A$.
If instead the results obtained with the full Wigner distribution in Fig.~\ref{fig:N_bar} did not reproduce this systematic trend, we would have concluded that this is implausible.
Electron scattering, since it can probe low-momentum states in the center of nuclei, provides an opportunity to rule out this model for finite systems.
We outline a number of possible approaches here.

In \emph{inclusive electron scattering}, $A(e,e')X$, low momentum states contribute in the region around the quasielastic peak, where Bjorken $x \approx 1$.
The nuclear matter data used in Ref.~\cite{Koch:2024qnz} is from the analysis of Ref.~\cite{Day:1989nw}, which uses the measurements of Whitney et al.~\cite{Whitney:1974hr}.
The other nuclear matter data are taken at higher energies, and don't show a clear quasielastic peak.
The Whitney measurements have significant uncertainties, compared for example to more recent data on ${}^{208}$Pb near the quasielastic peak~\cite{Zghiche:1993xg}.
A systematic study of more precise $(e,e')$ data for large $A$ near the quasielastic peak can be sensitive to the presence or absence of low-momentum suppression.
Theoretical benchmarks in this regime are challenging, one needs to take into account the effect of short- and long-range correlations, final-state interactions, two-body currents, and sizable Coulomb corrections.

A more direct way to probe low-momentum states is {\it single  nucleon knockout}, $A(e,e'p)X$.
In the plane-wave impulse approximation the distribution of the missing momentum $\mathbf{p}_m \equiv \mathbf{k}_e - \mathbf{k}_{e\prime} - \mathbf{k}_p$ measured in nucleon knockout from a state, is proportional to the momentum distribution of that state\footnote{The direct proportionality of the cross section as function of $p_m$ and the nuclear momentum distribution is lost if one goes beyond the plane-wave impulse approximation~\cite{Boffi93}.}.
Low-momentum nucleons are associated with $s$-states. Therefore, performing $(e,e'p)$ measurements at missing energy in the region where $s$-states would contribute would provide an opportunity to rule out the model presented here. To our knowledge, the only experiments that observed the lowest-lying $s$-state were performed on $^{12}$C. See Refs.~\cite{Lapikas:1999ss,PhysRevC.71.064610}.
However, in $^{12}$C a low momentum suppression might be absent, or too small to measure.
Indeed, the LDA and HFB predictions for $P_{f_q>1}$ drop quickly at small $A$ according to Fig.~\ref{fig:N_bar}, and HFB and RMF calculations give conflicting results in this regime. This indicates that the result for small $A$ is not robust. The missing momentum distribution associated with $s$-states have been measured for $^{40}$Ca and $^{208}$Pb~\cite{Lapikas:1993uwd, A1:2023cyw, A1:2023axb}, but only for the isolated levels at the top of the Fermi sea, for which the nuclear density is smallest. A modern-day effort for $(e,e'p)$ that probes low missing momentum, ideally in the region of the deeper-lying $s$-states, could rule out the theory presented here.

In $(e,e'p)$ experiments the cross section is suppressed at low $p_m$ because of a phase-space factor.
This kinematic suppression can be avoided in e.g. \emph{exclusive pion production} $A(e,e' p, \pi^-)X$.
In this case, for fixed scattering angles of the outgoing hadrons and fixed missing energy, the missing momentum is practically proportional to the kinetic energy of the proton (or pion).
This allows to measure the $p_m$ dependence without the kinematic zero, see e.g. Fig.~2 of Ref.~\cite{PhysRevC.48.816} where the missing momentum reaches a minimum $p_m\approx 10~\mathrm{MeV}$ for $T_p \approx 120~\mathrm{MeV}$. 
To our knowledge, the only data for exclusive single-pion production off nuclei sensitive to $s$-shell momentum distributions is for pion photoproduction off ${}^{12}$C~\cite{PiPhot}.

\section{Summary}
In Ref.~\cite{Koch:2024qnz}, the authors construct a model for nuclear matter that is composed of a momentum-shell of quarkyonic matter surrounded by a shell of nucleonic matter.
One of the salient features of this model is that low-momentum nucleon states are strongly depleted, leaving effectively a 'hole' in the nucleon momentum distribution.
This suppression of low momentum states was not ruled out by $(e,e^\prime)$ data for infinite nuclear matter~\cite{Day:1989nw}, as a matter of fact the data has the same feature of a  suppression of the cross section around the quasielastic peak.
This nuclear matter data is obtained from an extrapolation procedure of $(e,e')$ data for different nuclei to large mass $A$. Hence, if this suppression is a genuine feature, it should also be present in finite nuclei. Our main goal is to study if this is feasible using the same ingredients of Ref.~\cite{Koch:2024qnz}, by extending the analysis to finite-sized nuclei. 

To do so, the natural extension is to replace the occupation numbers in the Fermi sea by the nuclear Wigner distribution.
We use on the one hand the Wigner distribution assuming the local density approximation (LDA), a straightforward extension of~\cite{Koch:2024qnz}. And on the other hand the more realistic Wigner distribution obtained for spherical systems in mean field models.
We use relativistic mean field (RMF) and Hartree-Fock Bogoliubov (HFB) calculations to compute the nuclear Wigner distribution.
We point out some general features of the Wigner distribution, and show in particular that the distribution obtained in different independent-particle shell models (RMF, HF, and harmonic oscillator results of Ref.~\cite{Cosyn:2021ber}) are quantitatively similar.
This is understood by the limiting behavior of the Wigner distribution as $r \rightarrow 0$ or $k \rightarrow 0$. In these cases the contribution to the Wigner distribution from a filled shell is simply determined by the Fourier transform of the scalar momentum and position densities respectively. 
Then, since each shell contributes with a sign $(-1)^l$, the Wigner distribution up to intermediate $r,k$, is essentially determined by the angular momentum structure of the nuclear shell model and bulk properties such as the radii and Fermi momenta associated with different shells.
A code to compute the Wigner distributions is available upon request. They can play a role in semiclassical intranuclear cascade models which are crucial for the analysis of modern accelerator-based neutrino experiments~\cite{NUSTECWP}.

To study the influence of a possible quarkyonic regime, we convolute the nuclear Wigner distribution with the simple model of the quark momentum distribution of Ref.~\cite{Fujimoto:2023mzy}, to obtain the quark phase-space distribution $f_q$ in analogy with Ref.~\cite{Koch:2024qnz}.
In the LDA, identification of this phase-space distribution with occupation numbers in the (local) Fermi gas is straightforward. $f_q>1$ then naturally implies a violation of the quark Pauli principle. In the case of the full Wigner distribution this is not strictly necessary, and identifying an occupation number for quark states is not straightforward.
We show that even though the full Wigner distribution and the one obtained in the LDA behave completely differently, the quark phase-space distribution obtained from both are quantitatively similar. This is understood from the fact that the convolution, Eq.~(\ref{eq:F_q_rspace}), probes the one-body density matrix $\rho(\mathbf{r}+\mathbf{s},\mathbf{r}-\mathbf{s})$ only at small values $\mathbf{s} \lesssim R_N$, with $R_N$ the nucleon radius.
For small $s$, the one-body density matrix is approximated well in the LDA~\cite{Negele:1972zp}, see Fig.~\ref{fig:OBDM}.
The similarity between the quark phase space distribution from LDA and full calculation naturally becomes better as $A$ increases, since the LDA becomes an excellent approximation to the one-body density matrix up to large $s$ and large $r$. The results then naturally tend to the nuclear matter limit in the nuclear interior, see e.g. Fig.~\ref{fig:OBDM} for ${}^{208}$Pb.

We then compute a measure of the violation of the quark Pauli principle as the probability $P_{f_q>1}$. That is the percentage of baryons that violates the bound $f_q > 1$.
We study this probability for a set of 12 nuclei with $12 \leq A \leq 238$.
We find that this probability systematically tends to a fixed nuclear matter limit as function of $A$, for $A > 50$. One finds results compatible with the LDA with the full calculations, except for light nuclei, where the RMF and HFB results disagree.
We relate this probability to the momentum $k_1$ up to which the nucleon momentum distribution is suppressed in the model of Ref.~\cite{Koch:2024qnz}. We find that the ratio to the Fermi momentum $k_1/k_F$ slowly increases to the nuclear matter limit, with saturation for $A> 100$.
We thus conclude that, in this model, it is plausible that there exists suppression of low momentum states in finite nuclei, and moreover that the trend is systematic such that it can be seen after extrapolation to the nuclear matter limit~\cite{Day:1989nw}.

This is a counter-intuitive result, since it implies a direct influence of quarkyonic degrees of freedom on the nucleon momentum distribution. If this suppression is indeed present, one should be able to measure it in electron scattering facilities.
We give possible electron-nucleus scattering measurements on large nuclei that could rule out this model for finite nuclei.

\section*{Acknowledgments}
GAM  gratefully acknowledges the hospitality of  MIT LNS, UC Berkeley Physics Department, Lawrence Berkeley Lab and Argonne National Lab while this work was done. AN is supported by the Neutrino Theory Network under Award Number DEAC02-07CH1135. We  thank L. McLerran, N. Steinberg, A. Bulgac and O. Hen for useful discussions.

\bibliographystyle{apsrev4-1.bst}
\bibliography{bibliography}% Produces the bibliography via BibTeX.

\appendix

\section{Wigner distribution in the Relativistic Mean Field}
\label{app:Wigner_RMF}
In the relativistic mean field, the nuclear ground state is described as a Slater determinant of single-particle states with definite angular momentum and parity. They can be written as a four-component spinor
\begin{equation}
\psi_{n,\kappa}^{m_j}\left(\mathbf{r}\right) =  
\begin{pmatrix}
g_{n,\kappa}(r) \Phi_\kappa^{m_j}(\Omega_r) \\
f_{n,\kappa}(r) \Phi_{-\kappa}^{m_j}(\Omega_r).
\end{pmatrix},
\label{A1}
\end{equation}
We suppress the subscript denoting principal quantum number $n$ in the following for ease of notation.
Total and orbital angular momentum $J$ and $l$, are given by the relativistic angular momentum $\kappa$, $\lvert \kappa \rvert = J+1/2$, and $l = j  \pm 1/2$ for $\kappa = \pm \lvert \kappa \rvert$ respectively.
The spin spherical harmonics are 
\begin{equation}
    \Phi_{\kappa}^m\left(\Omega \right) = \sum_{s,m_l} ( l, m_l ; \frac{1}{2}, s | \left. J , m \right) Y_{l}^m\left(\Omega \right) \chi^{s},
\end{equation}
where $\chi^s$ are orthonormal spin-$1/2$ states,  $( l, m_l ; \frac{1}{2}, s | \left. J , m \right)$ are Clebsch-Gordan coefficients, and $Y_{l}^m$ are spherical harmonics.
The full one-body density matrix can be represented by a 4-by-4 matrix, but we only need to consider the trace for our purposes.
Indeed, for a filled shell we have
\begin{align}
\rho_\kappa(\mathbf{r}^\prime,\mathbf{r}) &= \sum_{m_J} \overline{\psi}_\kappa^{m_J}(\mathbf{r}^\prime) \gamma^0 \psi_\kappa^{m_J}(\mathbf{r}) \\
&= \mathrm{Tr}\left[\sum_{m_J} \psi_\kappa^{m_J} (\mathbf{r}) \left[ \psi_\kappa^{m_J}(\mathbf{r}^\prime) \right]^\dagger \right],
\end{align}
such that $\rho_\kappa(\mathbf{r},\mathbf{r})$ is the baryon density of the shell.
Other components of the matrix provide contributions proportional to e.g. the scalar or tensor density of the nucleus. 
The sum over $m_J$ can be performed, making use of the properties of the spin-spherical harmonics~\cite{Bechler_1993}, 
\begin{equation}
\label{eq:rho_kappa}
\rho_\kappa(\mathbf{r},\mathbf{r}^\prime) = \frac{2J+1}{4\pi}\left[ g_{\kappa}(r)  g_{\kappa}(r^\prime) P_{l}(x)
+ f_{\kappa}(r)  f_{\kappa}(r^\prime) P_{\overline{l}}(x) \right],
\end{equation}
where $P_l(x)$ are the Legendre polynomials, and $x = \cos\theta_{r,r^\prime} = \hat{\mathbf{r}}\cdot\hat{\mathbf{r^\prime}}$.
Here $\overline{l} = l + 1$ for $\kappa < 0$ and $\overline{l} = l-1$ for $\kappa > 0$.

The Wigner distribution is then given by 
\begin{equation}
W_\kappa(\mathbf{r}, \mathbf{p}) = \int \mathrm{d}\mathbf{s} e^{-i\mathbf{p}\cdot\mathbf{s}} \rho_\kappa( \mathbf{r} + \mathbf{s}/2, \mathbf{r} - s/2).
\end{equation}
Clearly, from Eq.~(\ref{eq:rho_kappa}), the one-body density depends on the relative angle $\cos\theta_{rs} = \hat{\mathbf{r}}\cdot\hat{\mathbf{s}}$, and thus $W(\mathbf{r}, \mathbf{p})$ depends on $\cos\theta_{pr} = \hat{\mathbf{r}}\cdot\hat{\mathbf{p}}$.
The explicit dependence is clear after integrating over the azimuthal angle
\begin{align}
&W(\mathbf{p}, \mathbf{r}) = 2\pi \int s^2\mathrm{d}s \mathrm{d}\cos\theta_{rs} J_0\left(ps\sin\theta_{pr}\sin\theta_{rs}\right) \times \nonumber \\
& \cos\left(ps \cos\theta_{pr} \cos\theta_{rs} \right) \rho_\kappa(\mathbf{r}-\mathbf{s}/2, \mathbf{r}+\mathbf{s}/2),
\label{eq:W_full}
\end{align}
with $J_0(x)$ the cylindrical Bessel function.
It is straightforward to compute the integral over $\cos\theta_{pr}$~\cite{NIST:DLMF_10_22_26}, so that the angle-averaged distribution can be computed as 
\begin{align}
\label{eq:angleint_wrp}
w_\kappa(r,p) &\equiv \frac{1}{2} \int \mathrm{d}\cos\theta_{pr} W(\mathbf{p},\mathbf{r}) \nonumber \\
&= 2 \pi \int s^2\mathrm{d}s \mathrm{d}\cos\theta~j_0\left(ps\right) \rho_\kappa(\mathbf{r}-\mathbf{s}/2, \mathbf{r}+\mathbf{s}/2),
\end{align}
with $j_0$ the spherical Bessel function.
The integrals (\ref{eq:W_full}), (\ref{eq:angleint_wrp}) are computed numerically using p-adaptive quadrature~\cite{cubature}. The size of the phase-space box $s \in [ 0 : s_{max}]$ is varied until convergence is reached.

\subsection{Properties of the Wigner distribution}
The distribution is normalized so that 
\begin{equation}
\int \frac{\mathrm{d}\mathbf{p}}{(2\pi)^3} W(\mathbf{r},\mathbf{p}) = \sum_{\kappa} \frac{2J_{\kappa}+1}{4\pi}\left[g_\kappa^2(r) + f_\kappa^2(r) \right] = \rho(\mathbf{r},\mathbf{r})
\end{equation}
is the baryon density.
The Wigner distribution has the same form for $r$ and $p$-space wavefunctions
\begin{equation}
\label{eq:WRP_pspace}
W_\kappa(\mathbf{r},\mathbf{p}) = \int \mathrm{d}\mathbf{s} e^{i\mathbf{r}\cdot\mathbf{s}} \rho_\kappa(\mathbf{p} - \mathbf{s}/2, \mathbf{p} + \mathbf{s}/2),
\end{equation}
where 
\begin{equation}
\rho_\kappa(\mathbf{p},\mathbf{p}^\prime) \equiv \sum_{m_j} \left[\phi^{m_j}_{\kappa}(\mathbf{p})\right]^\dagger\phi^{m_j}_\kappa(\mathbf{p}^\prime),
\end{equation}
and the momentum space wavefunction have the same structure as the $r$-space ones
\begin{equation}
\phi_\kappa^{m_j}(\mathbf{p}) = \int \mathrm{d}\mathbf{r} e^{-i\mathbf{r}\cdot\mathbf{p}} \psi_\kappa^{m_j}(\mathbf{r}) = 
\begin{pmatrix}
G_{\kappa}(p) \Phi_\kappa^{m_j}(\Omega_p) \\
F_{\kappa}(p) \Phi_{-\kappa}^{m_j}(\Omega_p)
\end{pmatrix},
\end{equation}
where the radial wavefunctions $G(p)$ and $F(p)$ are related to $g(r)$ and $f(r)$ by a Hankel transform.

The Wigner distribution will necessarily take on negative values in some part of the phase space.
It is bounded by the Cauchy-Schwarz inequality
\begin{align}
&\lvert W_\kappa^{m_J}(\mathbf{r},\mathbf{p}) \rvert = \lvert \int \mathrm{d}\mathbf{s} \left[\psi^{m_J}_\kappa(\mathbf{r} + \mathbf{s}/2)\right]^\dagger \psi_\kappa^{m_J}(\mathbf{r} - \mathbf{s}/2) e^{i\mathbf{p}\cdot\mathbf{s}} \rvert \nonumber \\
&\leq \int\mathrm{d}\mathbf{s} \left[ \psi_\kappa^{m_J}(\mathbf{r} + \mathbf{s}/2) \right]^\dagger \psi_\kappa^{m_J}(\mathbf{r} + \mathbf{s}/2) = 8
\end{align}
since the single-particle states are normalized to unity.
This bound then implies $\lvert W_\kappa(\mathbf{r},\mathbf{p}) \rvert \leq 8 (2J_\kappa+1)$ and $\lvert W(\mathbf{r},\mathbf{p})\rvert = \lvert \sum_\kappa W_\kappa(\mathbf{r},\mathbf{p} \rvert \leq 8A$.

For small $\mathbf{r}, \mathbf{p}$, $W_\kappa(\mathbf{r},\mathbf{p})$ comes close to saturating this bound.
Indeed, for $\mathbf{r} = 0$ we have
\begin{equation}
W_\kappa(0,\mathbf{p}) = 2^3 \int \mathrm{d} \mathbf{s} e^{i 2 \mathbf{p}\cdot\mathbf{s}} \rho_\kappa\left(\mathbf{s},-\mathbf{s}\right),
\end{equation}
and from Eq.~(\ref{eq:WRP_pspace}), one finds the equivalent expression for $\mathbf{p} = 0$.
From Eq.~(\ref{A1}), the RMF wavefunctions behave under parity as $\psi^{m_j}_{\kappa}(-\mathbf{r}) = (-1)^l \gamma^0 \psi^{m_j}_\kappa(\mathbf{r})$, 
  hence $\rho(\mathbf{x},-\mathbf{x})$ is proportional to the scalar density $\overline{\psi}\psi$, in both $r$ or $p$-space.
  With the scalar densities
  \begin{equation}
\
\rho^s_\kappa(x) = \sum_{m_j}\left[\psi_{\kappa}^{m_j}(\mathbf{x})\right]^\dagger\gamma^0\psi_\kappa^{m_j}(\mathbf{x}),
  \end{equation}
the $p$ dependence at $r=0$ and vice versa for each shell is given by
\begin{align}
\label{eq:wign_r0}
W_\kappa(0,\mathbf{p}) &=(-1)^{l} \left(2J + 1\right) 2^3 \int s^2 \mathrm{d}s j_0(2ps) \left[g_\kappa^2(s) - f_\kappa^2(s) \right] \nonumber \\
	&= (-1)^l 2^3 \mathcal{FT}\left\{ \rho^s_\kappa(r) \right\}\left( p/2 \right),
\end{align}
\begin{align}
\label{eq:wign_p0}
W_\kappa(\mathbf{r},0) &=(-1)^{l} \left(2J + 1\right) 2^3 \int s^2 \mathrm{d}s j_0(2rs) \left[G_\kappa^2(s) - F_\kappa^2(s) \right] \nonumber \\
	&= (-1)^l 2^3 \mathcal{FT}\left\{ \rho^s_\kappa(p) \right\}\left( r/2 \right).
\end{align}
That is, they are determined by the Fourier transform of the scalar density, and scalar momentum density respectively, with sign given by the angular momentum of the state.
Since the radial wavefunctions for the lower components are small ($f/g \sim \hat{\vb{p}}/2M$), $\mathcal{FT}\left\{ \rho^s_\kappa(\mathbf{x}) \right\}\left( 0 \right) \approx (2J_\kappa+1)$, and hence the Wigner distribution for each shell almost saturates the bound.
\subsection{Non relativistic wavefunctions}
The discussion above remains practically unchanged when non-relativistic wavefunctions are considered.
In this case, one can write the wavefunction as the two component wavefunction
\begin{equation}
\label{eq:nr_wf}
\psi_{n,\kappa}^{m_j} = u_{n,\kappa}(r) \Phi_\kappa^{m_j}\left(\Omega_r\right).
\end{equation}
All results are then the same as in the RMF case, one simply drops the lower component wavefunctions $f(r)$.
For nuclei with partially filled $(n,\kappa)$ shells, we use the canonical states $\psi^{m_j}_{n,\kappa}$ and fractional occupation numbers $v_{n,\kappa}^2$ obtained from the HFB calculation.
The canonical states are defined as the states that diagonalize the one-body density matrix
\begin{equation}
\rho(\mathbf{r},\mathbf{r}^\prime) = \sum_{n,\kappa,m_j} v_{n,\kappa}^2\left[\psi^{m_j}_{n,\kappa}(\mathbf{r})\right]^\dagger\psi^{m_j}_{n,\kappa}(\mathbf{r^\prime}).
\end{equation}
Since spherical symmetry is imposed in HFBRAD~\cite{Bennaceur:2005mx}, they have the form of Eq.~(\ref{eq:nr_wf}). 
We make use of the completeness relation of Eq.~(\ref{eq:rho_kappa}), we simply include the fractional occupation number $v_{n,\kappa}^2 \leq 1$ in the contribution of the shell. 
The bounds derived on the Wigner distribution remains the same, for each shell $\lvert W_\kappa(\mathbf{r},\mathbf{p})\rvert \leq 2^3(2J_\kappa+1)$.
In this case however, the scalar and vector densities become equal, such that $W_\kappa(0,0) = v_{\kappa}^2 2^3(-1)^l(2J_\kappa+1)$. Hence the Wigner distribution exactly saturates the bound for a filled shell where $v^2 = 1$.

We use surface pairing in the HFB calculations. For the set of nuclei we consider, pairing is small, and the $v_{n,\kappa}^2$ are similar to those found in a HF calculation, with minimal strength in higher-lying canonical states. The results are not sensitive to approximating the HFB results with shell occupations based on a pure mean field, the choice of pairing or functional would not influence our results significantly.

\onecolumngrid
\newpage

\end{document}